\documentclass[aps,prl,twocolumn,nofootinbib, superscriptaddress]{revtex4-2}

\usepackage{graphicx}
\usepackage{epstopdf}
\usepackage{amsmath}
\usepackage{amssymb}
\usepackage{mathrsfs}
\usepackage{amsthm}
\usepackage{bm}
\usepackage{url}
\usepackage[T1]{fontenc}
\usepackage{csquotes}
\MakeOuterQuote{"}
\usepackage[braket]{qcircuit}
\usepackage{color}
\usepackage[ruled,lined]{algorithm2e}
\usepackage{algorithmic}
\usepackage{multirow}
\usepackage{makecell}

\newtheoremstyle{note}
  {\topsep/2}               
  {\topsep/2}               
  {}                      
  {\parindent}            
  {\itshape}              
  {.}                     
  {5pt plus 1pt minus 1pt}
  {}

\theoremstyle{note}

\theoremstyle{definition}

\theoremstyle{remark}

\newcommand{\tr}{\operatorname{tr}}

 \newcommand{\rmd}{\mathrm{d}}
 \newcommand{\rme}{\mathrm{e}}
 \newcommand{\rmi}{\mathrm{i}}

 \newcommand{\caH}{\mathcal{H}}

 \newcommand{\caQ}{\mathcal{Q}}

 \newcommand{\id}{1}

 \newcommand{\scrA}{\mathscr{A}}
 \newcommand{\scrB}{\mathscr{B}}
 \newcommand{\scrC}{\mathscr{C}}

\newcommand{\be}{\begin{equation}}
\newcommand{\ee}{\end{equation}}
\newcommand{\ba}{\begin{align}}
\newcommand{\ea}{\end{align}}

\def\<{\langle}  
\def\>{\rangle}  

\def\eqref#1{\textup{(\ref{#1})}}  
\newcommand{\eref}[1]{Eq.~\textup{(\ref{#1})}}

\newcommand{\fref}[1]{Fig.~\ref{#1}}
\newcommand{\Fref}[1]{Figure~\ref{#1}}

\newcommand{\tref}[1]{Table~\ref{#1}}

\newcommand{\cref}[1]{Conjecture~\ref{#1}}
\newcommand{\Cref}[1]{Conjecture~\ref{#1}}

\newcommand{\rcite}[1]{Ref.~\cite{#1}}
\newcommand{\rscite}[1]{Refs.~\cite{#1}}

\begin{document}
\title{Experimental Demonstration of  Inequivalent Mutually Unbiased Bases}

	\author{Wen-Zhe Yan} 
    \thanks{These authors contributed equally to this work.}
	\affiliation{CAS Key Laboratory of Quantum Information, University of Science and Technology of China, Hefei 230026, People's Republic of China}
	\affiliation{CAS Center For Excellence in Quantum Information and Quantum Physics, University of Science and Technology of China, Hefei 230026, People's Republic of China}
 
    \author{Yunting Li}
    \thanks{These authors contributed equally to this work.}
    \affiliation{State Key Laboratory of Surface Physics and Department of Physics, Fudan University, Shanghai 200433, China}
    \affiliation{Institute for Nanoelectronic Devices and Quantum Computing, Fudan University, Shanghai 200433, China}
    \affiliation{Center for Field Theory and Particle Physics, Fudan University, Shanghai 200433, China}
    
	\author{Zhibo Hou} 
	\email{houzhibo@ustc.edu.cn}
	\affiliation{CAS Key Laboratory of Quantum Information, University of Science and Technology of China, Hefei 230026, People's Republic of China}
	\affiliation{CAS Center For Excellence in Quantum Information and Quantum Physics, University of Science and Technology of China, Hefei 230026, People's Republic of China}
 
    \author{Huangjun Zhu}
    \email{zhuhuangjun@fudan.edu.cn}
    \affiliation{State Key Laboratory of Surface Physics and Department of Physics, Fudan University, Shanghai 200433, China}
    \affiliation{Institute for Nanoelectronic Devices and Quantum Computing, Fudan University, Shanghai 200433, China}
    \affiliation{Center for Field Theory and Particle Physics, Fudan University, Shanghai 200433, China}
    
	\author{Guo-Yong Xiang}
	\email{gyxiang@ustc.edu.cn}
	\affiliation{CAS Key Laboratory of Quantum Information, University of Science and Technology of China, Hefei 230026, People's Republic of China}
	\affiliation{CAS Center For Excellence in Quantum Information and Quantum Physics, University of Science and Technology of China, Hefei 230026, People's Republic of China}
 
	\author{Chuan-Feng Li}
	\affiliation{CAS Key Laboratory of Quantum Information, University of Science and Technology of China, Hefei 230026, People's Republic of China}
	\affiliation{CAS Center For Excellence in Quantum Information and Quantum Physics, University of Science and Technology of China, Hefei 230026, People's Republic of China}
 
	\author{Guang-Can Guo}
	\affiliation{CAS Key Laboratory of Quantum Information, University of Science and Technology of China, Hefei 230026, People's Republic of China}
	\affiliation{CAS Center For Excellence in Quantum Information and Quantum Physics, University of Science and Technology of China, Hefei 230026, People's Republic of China}

\begin{abstract}
Quantum measurements based on mutually unbiased bases (MUB) play  crucial roles
in foundational studies and quantum information processing. It is known that there exist inequivalent MUB, but little is known about their operational distinctions, not to say experimental demonstration. In this work, by virtue of a simple estimation problem we experimentally demonstrate the operational distinctions between inequivalent triples of MUB in dimension 4 based on high-precision photonic systems. The experimental estimation fidelities coincide well with the theoretical predictions with only 0.16\% average deviation, which is 25 times less than the  difference (4.1\%) between  the maximum estimation fidelity and the minimum estimation fidelity. Our experiments clearly demonstrate that inequivalent MUB have  different information extraction capabilities and different merits for quantum information processing.

\end{abstract}

\date{\today}

\maketitle

\textit{Introduction}.---Quantum measurements play a key role in extracting information from quantum systems and in achieving various quantum information processing tasks, such as quantum computation, quantum communication, quantum metrology, quantum sensing, and quantum simulation \cite{vonNeumann2018,nielsen_chuang_2010,Blundell2017}. Rank-1 projective measurements are the simplest quantum measurements discussed in most elementary textbooks on quantum mechanics. Nevertheless, 
their properties become elusive if we consider two or more projective measurements. Since each rank-1 projective measurement is tied to an orthonormal basis, and vice versa, the study of rank-1 projective measurements is intertwined with the study of orthonormal bases.

Two rank-1 projective measurements are \emph{mutually unbiased} or complementary if the outcome of one measurement is completely random whenever the outcome of the other measurement is certain. The corresponding bases are called \emph{mutually unbiased bases} (MUB) \cite{Schwinger1960,Ivonovic1981,WOOTTERS1989363,DURT2010}. 
MUB are closely tied to the complementarity principle \cite{BOHR1928} and uncertainty relations \cite{Heisenberg1927,Robertson1929,Busch2014,Wehner_2010,Coles2017},
which play  key roles in quantum mechanics. Moreover, MUB have found numerous applications in quantum information processing, including quantum cryptography \cite{Mayers1998, DURT2010,Coles2017,Armin2021}, quantum
random access codes \cite{Aguilar2018, Farkas2019}, quantum state estimation \cite{Ivonovic1981,WOOTTERS1989363,Roy2007,Zhu2014,Adamson2010}, and quantum verification \cite{LiBipartite2019, ZhuMES2019}.

Two sets of MUB are  (unitarily) \emph{equivalent} if they can be turned into each other by unitary transformations up to the order of basis elements and overall phase factors. Otherwise, they are \emph{inequivalent}. 
Inequivalent MUB exist in certain dimensions at least four \cite{Brierley2009AllMU}, and this intriguing phenomenon has attracted the attention of many researchers in various research areas \cite{DURT2010,Brierley2009AllMU,Kantor2012,Designolle2019,Aguilar2018,Farkas2019,Hiesmayr_2021, Armin2021,Zhu2022QMQSE}. However, little is known about the operational distinctions between inequivalent MUB, and there is no experimental demonstration before as far as we know.

As notable exceptions,  Aguilar et al. showed that inequivalent MUB can achieve different success probabilities in quantum
random access codes \cite{Aguilar2018}. 
Designolle et al.  showed that inequivalent MUB may have different degrees of measurement incompatibility as quantified by the noise robustness \cite{Designolle2019}. Hiesmayr et al.  showed  that some MUB are more effective than others in detecting entanglement \cite{Hiesmayr_2021}. It is not clear whether these results can be demonstrated in experiments in the near future. Very recently, starting from a simple estimation problem, one of the authors showed that inequivalent MUB may have different information-extraction capabilities and can be distinguished by the \emph{estimation fidelity} \cite{Zhu2022QMQSE}, which is amenable to experimental demonstration.

In this work, using photonic systems we experimentally demonstrate the operational distinctions between inequivalent triples of MUB in dimension 4 based on a simple three-copy estimation problem.
To this end, we use polarization and path degrees of freedom of a photon to form a ququad. Then we  implement a three-copy estimation protocol in which the projective measurements are determined by triples of MUB, so that the estimation fidelities are tied to the intrinsic properties of MUB. The projective measurements we realized have  average fidelity above 0.995. The experimental estimation fidelities coincide well with the theoretical predictions with only 0.16\% average deviation, which is 25 times less than the  difference (4.1\%) between  the maximum estimation fidelity and the minimum estimation fidelity. In this way, our experiments clearly demonstrate different information extraction capabilities of inequivalent MUB, which has never been demonstrated before.

\textit{A simple estimation problem}.---Suppose a quantum device can prepare  $N$ copies of a  random pure quantum state $\rho$ on a $d$-dimensional Hilbert space $\caH$ according to the Haar measure. We are asked to estimate the identity of $\rho$ as accurately as possible as quantified by the average fidelity. If we perform the quantum measurement characterized by the positive operator-valued measure (POVM) $\scrA = \{A_j\}_j$ on $\rho^{\otimes N}$, then the probability of obtaining outcome $A_j$ is $p_j = \tr(\rho^{\otimes N} A_j)$. Let $\hat{\rho}_j$ be the estimator corresponding to outcome $j$. Let $P_{N+1}$ be the projector onto the symmetric subspace in $\caH^{\otimes (N+1)}$ and $D_{N+1}=\tr(P_{N+1})$. 
Then the average fidelity reads \cite{Zhu2022QMQSE}
\begin{align}
  \bar{F} &= \sum_j \int \rmd \rho p_j \tr(\rho \hat{\rho}_j)
  = \frac{\sum_j \tr [{Q}(A_j) \hat{\rho}_j) ]}{(N+1)! D_{N+1}}  \nonumber \\
  & \le F(\scrA):= \sum_j \frac{\lVert {Q}(A_j) \rVert }{(N+1)! D_{N+1}} ,  \label{eq:average_F}
\end{align}
where the integration is over the set of all pure states with Haar measure and
\begin{equation}
    {Q}(A_j) := (N+1)!\tr_{1,\dots,N} [P_{N+1} (A_j \otimes \id)].
\end{equation}
The upper bound in \eref{eq:average_F} is saturated iff each estimator $\hat{\rho}_j$ is supported in the eigenspace of ${Q}(A_j) $ associated with the largest eigenvalue.
Here  $F(\scrA)$  is the maximum average fidelity that can be achieved by the POVM $\scrA$ and is called the \emph{estimation fidelity} \cite{Zhu2022QMQSE}. It encodes valuable information about the POVM and is pretty useful for understanding elementary quantum measurements.

In the above analysis, the ensemble of Haar random pure states can be replaced by any ensemble of pure states that forms a \emph{t-design} with $t = N+1$,
which might be more appealing to practical applications.
Recall that a set of $K$ states $\{|\psi_j \>\}_j$ in  $\caH$ is a  $t$-design if $\sum_j (|\psi_j\>\<\psi_j|)^{\otimes t} $ is proportional to the projector $P_t$ onto the symmetric subspace in $\caH^{\otimes t}$ \cite{Renes2004, Zauner2011, Scott2006} or, equivalently, if 
the $t$th \emph{frame potential} 
\begin{equation}
    \Phi_t(\{|\psi_j \>\}_j): = \frac{1}{K^2} \sum_{j,k}| \< \psi_j|\psi_k\>|^{2t}
\end{equation}
saturates the inequality $\Phi_t(\{|\psi_j \>\}_j) \ge 1/D_t$.

\begin{figure}[t]
	\centering	
	\includegraphics[width=0.49\textwidth]{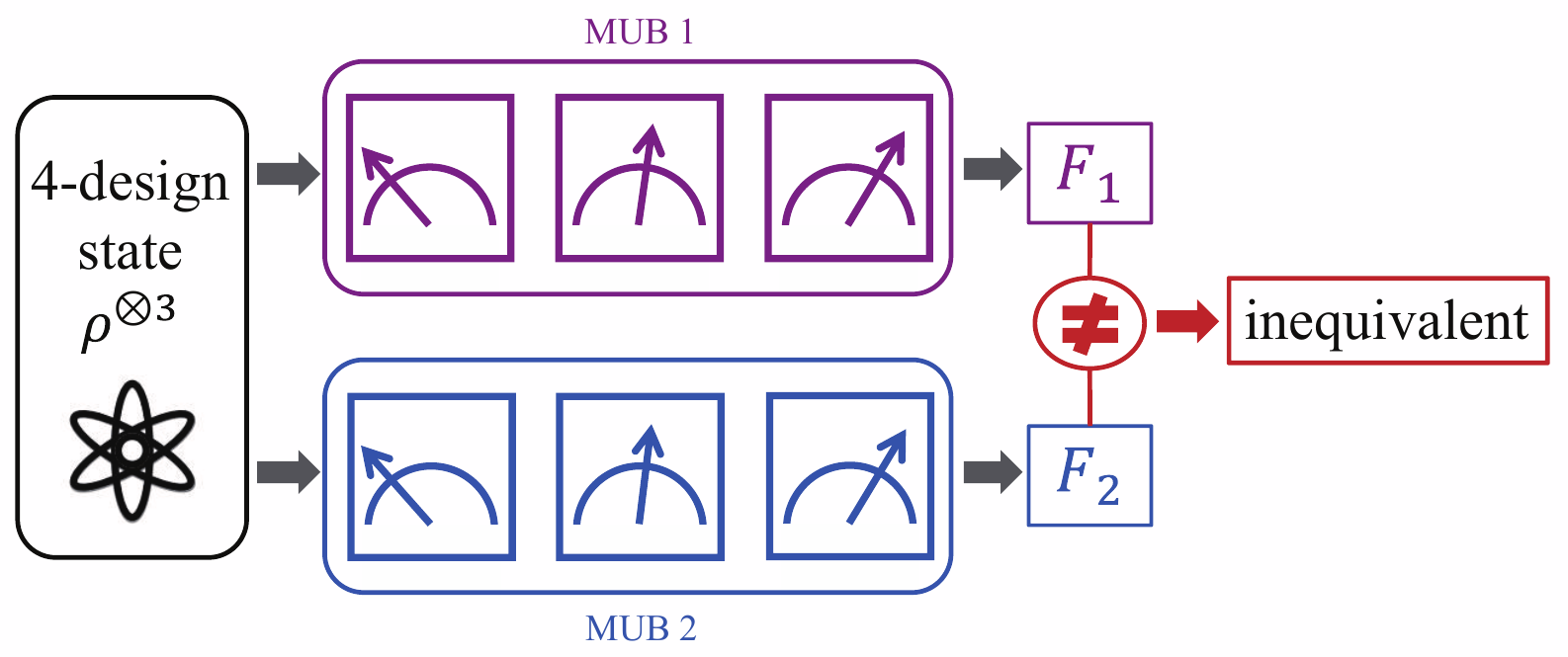}
	\caption{\label{fig:framework} The basic idea for distinguishing inequivalent tripes of  MUB. After preparing three copies of each state $\rho$ in a given 4-design  and performing three projective measurements associated with each triple of MUB,  the fidelity between the optimal estimator $\hat{\rho}$ and $\rho$ is evaluated. The estimation fidelity of each MUB is determined by averaging the fidelity $\tr(\rho\hat{\rho})$ over the 4-design and many repetitions. Two MUB are  inequivalent if their estimation fidelities are different. 
	} 
\end{figure}

\textit{Distinguishing inequivalent MUB}.---Here we are particularly  interested in inequivalent triples of MUB in dimension 4, which can be distinguished by the three-copy estimation fidelity  \cite{Zhu2022QMQSE} as illustrated in \fref{fig:framework}. The three bases are denoted by $\{|\alpha_j\> \}_j$, $\{|\beta_j\> \}_j$,  and $\{|\gamma_j\> \}_j$, respectively, where $j=1,2,3,4$; the corresponding rank-1 projective measurements read
$\scrA = \{|\alpha_j\> \< \alpha_j| \}_j$, $\scrB = \{|\beta_j\> \< \beta_j| \}_j$, and $\scrC = \{|\gamma_j\> \< \gamma_j| \}_j$. To be specific, the first basis  is chosen as the computational basis; the second and third bases correspond to the columns of the two Hadamard matrices, respectively \cite{Brierley2009AllMU,Zhu2022QMQSE}:
\begin{equation}\label{eq:POVM_B&C}
\begin{aligned}
  H_\scrB &= 
  \frac{1}{2}
  \begin{pmatrix}
    1 & 1 & 1 & 1 \\
    1 & \rmi \rme^{\rmi x} & -1 & -\rmi \rme^{\rmi x} \\
    1 &  -1 & 1 & -1 \\
    1 &  -\rmi \rme^{\rmi x} & -1 & \rmi \rme^{\rmi x}
  \end{pmatrix},\\
  H_\scrC &= 
  \frac{1}{2}
  \begin{pmatrix}
    1 & 1 & 1 & 1 \\
    -\rme^{\rmi y} & \rme^{\rmi z} & \rme^{\rmi y} & -\rme^{\rmi z} \\
    1 & -1 & 1 & -1 \\
    \rme^{\rmi y} & \rme^{\rmi z} & -\rme^{\rmi y} & -\rme^{\rmi z}
  \end{pmatrix},
\end{aligned}
\end{equation}
where $x,y,z \in [0,\pi]$ are three real parameters. By construction, it is easy to verify that
\begin{align}
|\<\alpha_j|\beta_k\>|^2=|\<\beta_k|\gamma_l\>|^2=|\<\alpha_j|\gamma_l\>|^2=\frac{1}{4}
\end{align}
for $j,k,l=1,2,3,4$. So the three bases for given $x,y,z$ are indeed mutually unbiased. As shown in \rcite{Zhu2022QMQSE},  the estimation fidelity $ F_{\mathrm{MUB}}(x,y,z):=F(\scrA\otimes\scrB\otimes \scrC)$
can be used to distinguish inequivalent MUB. Moreover,
the difference between the maximum and minimum estimation fidelities  is about 4.1\%, which is amenable to experimental demonstration.

\begin{figure*}[htbp]
	\centering	
	\includegraphics[width=0.8\textwidth]{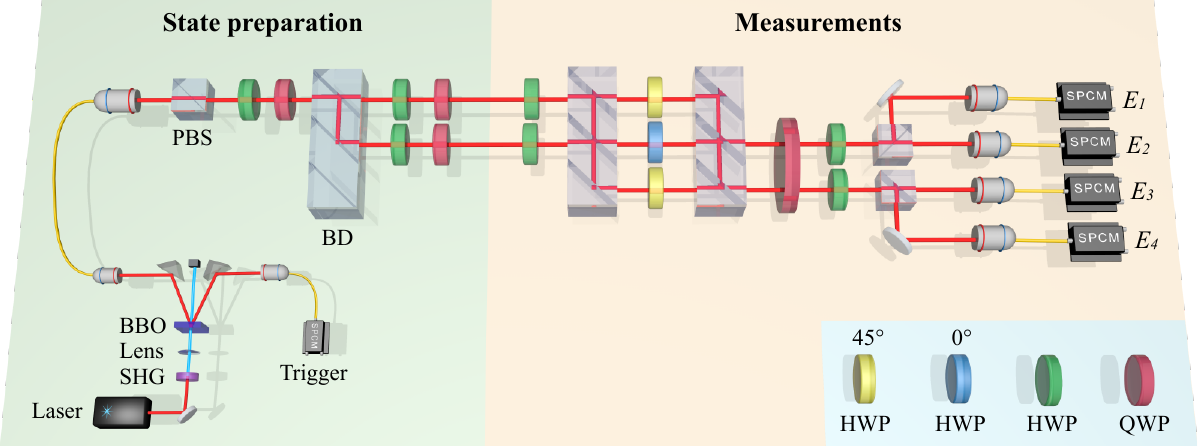}
	\caption{\label{fig:expsetup} Experimental setup. 
The module of state preparation generates  a single photon using the type-II phase-matched spontaneous parametric down-conversion (SPDC) process and prepares  the desired ququad state in polarization and path degrees of freedom. The measurement module  implements one of the three projective measurements $\scrA$, $\scrB$, and $\scrC$ associated with a triple of MUB. The four detectors at positions $E_1$ to $E_4$ correspond to the four outcomes of the measurement. Key elements include PBS (polarizing beam splitter), HWP (half wave plate), QWP (quarter wave plate), and BD (beam displacer). The wave plates which have not been marked with specific angles are the ones that need to be rotated during the experiments.
    } 
\end{figure*}

To determine the estimation fidelity in experiments,  we prepare three copies of each ququad state $\rho_i$  from a given 4-design composed of $K$ states  and perform the three projective measurements $\scrA, \scrB$, and $\scrC$, respectively.
If the outcomes $j,k,l$ are obtained, then we construct an estimator $\hat{\rho}_{jkl}$  on the eigenspace of $ \caQ(|\alpha_j\> \< \alpha_j| \otimes |\beta_k\> \< \beta_k| \otimes |\gamma_l\> \< \gamma_l|)$ corresponding to the largest eigenvalue and evaluate its fidelity with $\rho_i$. 
To suppress statistical fluctuation,  the same measurements are repeated $M$ times for each state $\rho_i$ in the 4-design. Denote by $\hat{\rho}_{im}$ the $m$th estimator for the $i$th state $\rho_i$, where $m=1,2,\dots, M$ and $i=1,2,\dots, K$. Then the estimation fidelity  is calculated as follows:
\begin{equation}\label{eq:F_MUB}
    F_{\mathrm{MUB}}(x,y,z) = \frac{1}{KM} \sum_i \sum_m \tr(\rho_i \hat{\rho}_{im}).
\end{equation}

To facilitate experimental realization, we constructed two 4-designs  with small cardinalities in dimension 4. The first 4-design is constructed from an orbit of length 960 of the restricted Clifford group and is referred to as the \emph{Clifford 4-design} henceforth. By  contrast,  the full Clifford orbit has length 3840 (see Secs.~S1 and S2 in the Supplemental Material \cite{supp}). This theoretical result is of independent interest, given that the Clifford group is only a 3-design, and the restricted Clifford group is only a 2-design \cite{Gottesman1997,Zhu2017_3design,Webb2016,zhu2016clifford}. 
The second one is a (approximate) \emph{numerical 4-design}  composed of 200 states, which is generated using the optimization algorithm  in \rscite{HUGHES202184,tightframes} (see Sec.~S3 in the Supplemental Material \cite{supp}).

\textit{Experimental setup}.---The experimental setup for realizing the three-copy estimation protocol is illustrated in \fref{fig:expsetup}.  We use the polarization (horizontal H and vertical V) and the path (up and down) degrees of freedom of a single photon to form a ququad. The setup is composed of two modules: the state preparation module, which can generate an arbitrary ququad state, and the measurement module, which can perform one of the projective measurements $\scrA$, $\scrB$, and $\scrC$.

In the state preparation process, a light pulse with a central wavelength of 780 nm first passes through a frequency doubler. Then the ultraviolet pulse is focused onto a BBO crystal cut for
the type-II phase-matched spontaneous parametric down-conversion (SPDC) process to create a pair of photons. One
photon is detected by a single-photon-counting module (SPCM) 
as a trigger, while the other acts as a heralded single-photon source.
The single photon is initialized in H polarization by a polarizing beam splitter (PBS).
Any polarization state can be generated by a combination of a half-wave plate (HWP) and a quarter-wave plate (QWP). A beam displacer (BD) which separates the H component and V component by 4 mm transforms this polarization state into a path state. The following two combinations of a HWP and a QWP adjust the polarization states in the two paths so as to generate the desired ququad state. Then, the state is sent into the measurement module, which performs one of the projective measurements $\scrA$, $\scrB$, and $\scrC$ (see Sec.~S4 in the Supplemental Material \cite{supp}). The regulation of the parameters $x,y,z$ featuring in  $\scrB$ and $\scrC$   is realized by changing the rotation angles of some HWPs.  Four SPCMs at the end correspond to four outcomes of the measurement.

\textit{Experimental results.---}In our experiment, we considered 18 triples of MUB corresponding to the parameters $x=\pi/2$, $y\in\{0, \pi/2\}$, and $z\in\{0, \pi/8, \pi/4, ..., \pi\}$, which share the two bases $\scrA$ and $\scrB(x=\pi/2)$. To characterize each projective measurement that was actually realized, we sent 36 states, the tensor products of the six eigenstates of three Pauli operators, to the measurement device and performed quantum measurement tomography. Each state was prepared and measured 10000 times. Then the four projectors were reconstructed from the measurement statistics using the method in \rcite{Fiur01maximum}
and the overall fidelity was evaluated as in \rcite{hou2018deterministic}. 
This procedure was repeated 10 times to determine the mean fidelity and error bar (standard deviation). Overall fidelities  of the realized measurements for $\scrA$ and $\scrB(x=\pi/2)$ are 0.9990$\pm$0.0001 and 0.9977$\pm$0.0003, respectively, while those for $\scrC(y,z)$ are shown in \tref{tab:tomography}. 
The average overall fidelity of these measurements is above 0.995, demonstrating that they were realized with high quality.

\begin{table*}[tbp]
        \renewcommand{\arraystretch}{1.3}
	\centering
	\caption{\label{tab:tomography} Overall fidelity of the measurement $\scrC(y,z)$ realized  in the experiment. Each fidelity value is the average over 10 repeated reconstructions. The error bar indicates the standard deviation of 10 repetitions.}
          \begin{tabular*}{1\textwidth}{@{\extracolsep{\fill}} c c c c c c c c c c}
		\hline \hline
		$z$  &  0 & $\pi/8$ & $\pi/4$ & $3\pi/8$ & $\pi/2$ & $5\pi/8$ & $3\pi/4$ & $7\pi/8$ & $\pi$\\
		\hline
		$\scrC(y=\pi/2)$ & 0.9979(2) & 0.9966(4) & 0.9943(12) & 0.9962(8) & 0.9975(8) & 0.9966(10) & 0.9948(8) & 0.9956(7) & 0.9975(4)\\
		$\scrC(y=0)$ & 0.9977(3) & 0.9965(7) & 0.9938(10) & 0.9959(10) & 0.9976(2) & 0.9954(12) & 0.9939(7) & 0.9947(10) & 0.9978(4)\\
		\hline \hline
	\end{tabular*}
\end{table*}

\begin{figure}[tbp]
	\centering	
	\includegraphics[width=0.8\linewidth]{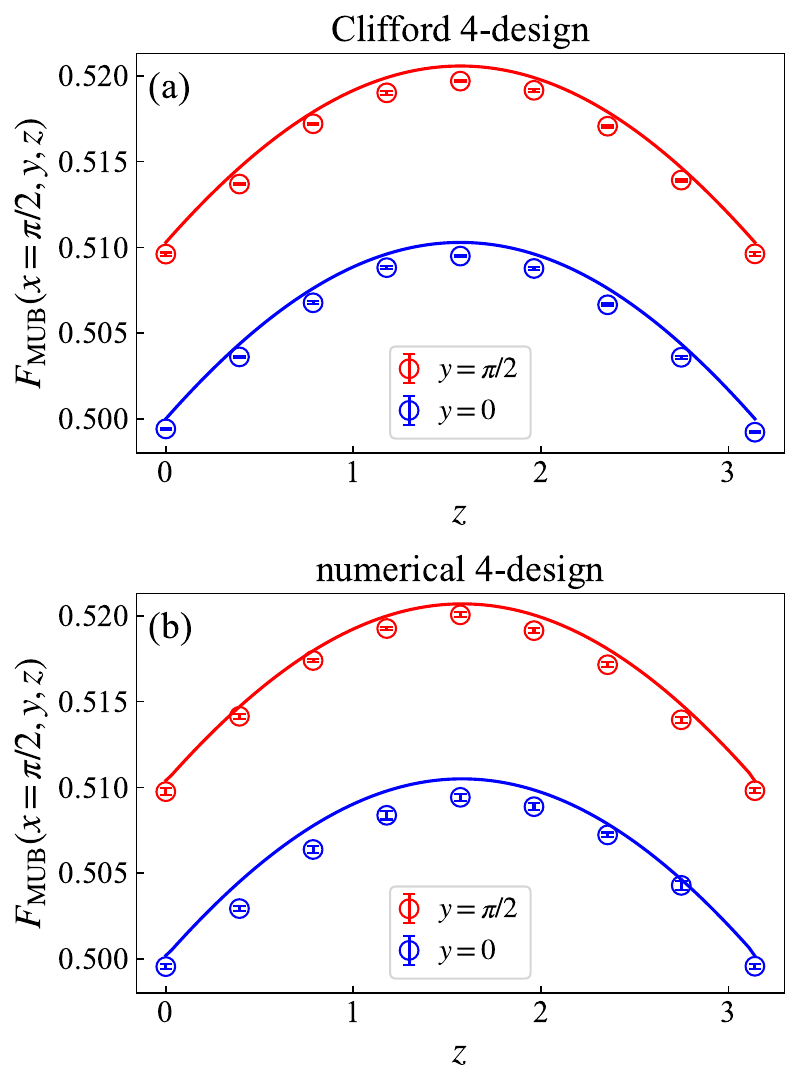}
	\caption{\label{fig:expresults} Experimental results (circles with error bars) and theoretical predictions (solid lines) on the three-copy estimation fidelity $F_\mathrm{MUB}(x,y,z)=F(\scrA \otimes \scrB \otimes \scrC)$ based on the  Clifford 4-design (a) and numerical 4-design (b), respectively. The error bars indicate the standard deviations of 10  repeated experiments.} 
\end{figure}

\begin{figure}[tbp]
	\centering	
\includegraphics[width=0.8\linewidth]{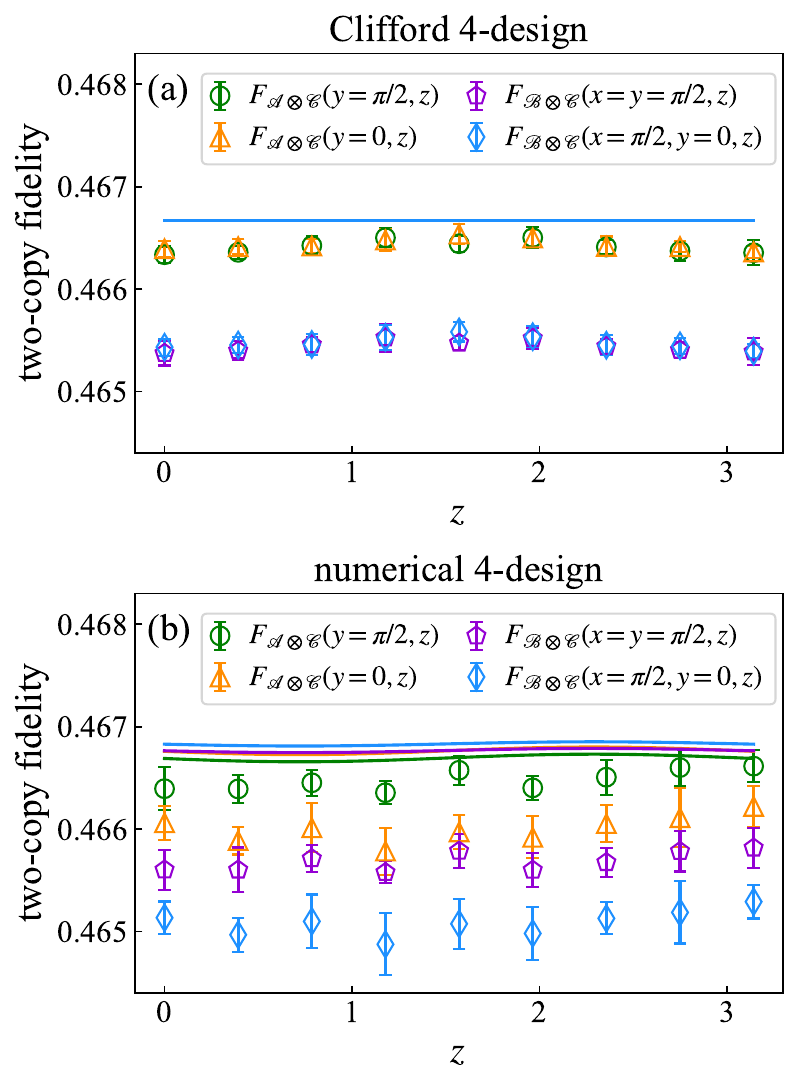}
	\caption{\label{fig:two copy} Experimental results (markers with error bars) and theoretical predictions (solid lines) on the two-copy estimation fidelities achieved by the product measurements $\scrA \otimes \scrC$ and $\scrB \otimes \scrC$ based on the Clifford 4-design (a) and numerical 4-design (b), respectively. The error bars indicate the standard deviations of 10 repeated experiments.} 
\end{figure}

\begin{table}[tbp]
        \renewcommand{\arraystretch}{1.3}
	\centering
	\caption{\label{tab:deviation}Average and maximum deviations between experimental (three-copy and two-copy) estimation fidelities and theoretical predictions based on the Clifford 4-design and numerical 4-design, respectively.}
	\begin{tabular*}{1\linewidth}{@{\extracolsep{\fill}} c c c c}
		\hline \hline 
		  &4-design   &  average &  maximum \\
		\hline
  
		\multirow{2}{*}{three-copy}
		&Clifford &0.0008 & 0.0010\\
		&numerical & 0.0008& 0.0016\\
  
		\multirow{2}{*}{two-copy}
		&Clifford & 0.0007& 0.0013\\

		&numerical & 0.0009 & 0.0019\\
		\hline \hline 
	\end{tabular*}
\end{table}

Next, we implemented the three-copy estimation protocol  to determine the estimation fidelity $F_\mathrm{MUB}(x,y,z)$.  To this end, we prepared three copies of each state in the Clifford 4-design and performed the three projective measurements $\scrA$, $\scrB(x=\pi/2)$, and $\scrC(y,z)$, respectively. To suppress statistical fluctuation and determine the error bar, the preparation and measurement procedure were repeated  $10000\times10$ times. For simplicity, we share the measurement outcomes of $\scrA$ and $\scrB(x=\pi/2)$ for all 18 sets of MUB. The estimation fidelity $F_\mathrm{MUB}(x,y,z)$ calculated by \eref{eq:F_MUB} are shown in plot (a) in \fref{fig:expresults}. The experimental results (circles with error bars) coincide well with the theoretical predictions (solid lines). This claim is further corroborated by \tref{tab:deviation}, which shows the average and maximum  deviations 
between experiments and theory. 
The experimental errors mainly come from the instability and drift of the phases of the Mach-Zehnder interferometers.
\Fref{fig:expresults} clearly delineates the variation of the estimation fidelity with the parameters $y,z$ for $x=\pi/2$, 
which highlights the operational distinction between inequivalent MUB.  Notably, the estimation fidelity reaches the maximum 0.5197  at $x=y=z=\pi/2$ and  the minimum 0.4992  at $x=\pi/2$, $y=0$, $z=\pi$; the difference  0.0205 is 25 times larger than the average deviation  shown in \tref{tab:deviation}.

Next, we implemented the three-copy estimation protocol based on the numerical 4-design instead of the Clifford 4-design. The results shown in plot (b) in \fref{fig:expresults} and \tref{tab:deviation} are quite similar to the counterparts based on the Clifford 4-design, although the two 4-designs are very different. These results  further demonstrate that the operational distinction between inequivalent MUB is independent of the choice of   4-designs. Incidentally, inequivalent triples of MUB in
dimension 4 cannot be distinguished by noise robustness considered in \rcite{Designolle2019}, which manifests the advantage of our approach.

As  comparison,   the two-copy estimation fidelity achieved by any product measurement based on MUB  equals 0.4667 \cite{Zhu2022QMQSE}, assuming that the state ensemble forms an ideal 4-design. Note that inequivalent MUB cannot be distinguished by the two-copy estimation fidelity, which provides information only about pairwise overlaps of the basis states \cite{Zhu2022QMQSE}. To demonstrate this result,
we reprocessed the experimental data to determine the two-copy estimation fidelity. 
The experimental two-copy estimation fidelities achieved by $\scrA \otimes \scrB(x=\pi/2)$ are 0.4664 $\pm$ 0.0001 and 0.4668 $\pm$ 0.0001 based on the Clifford 4-design and numerical 4-design, respectively, while those for $\scrA \otimes \scrC$ and $\scrB \otimes \scrC$ are shown in \fref{fig:two copy} together with theoretical predictions. 
The average and maximum deviations  are  shown in \tref{tab:deviation}. Again, the experimental results agree very well with theoretical predictions even if the numerical 4-design is not ideal.

\textit{Summary}.---In this work, we implemented a three-copy estimation protocol to demonstrate the operational distinction between inequivalent MUB. In our experiments, we used polarization and path degrees of freedom of a photon to form a ququad and performed projective measurements associated with 18 triples of MUB in dimemnsion 4 with high quality. The experimental estimation fidelities agree  well with  theoretical predictions with only 0.16\% average deviation, which is accurate enough to distinguish inequivalent MUB. Our 
experiments clearly demonstrate that inequivalent MUB may have different information extraction capabilities, which have operational consequences. These results are of intrinsic interest not only to foundational studies, but also to many tasks in quantum information processing, such as quantum state estimation, entanglement detection, and quantum communication.

\begin{acknowledgements}
The work at the University of Science and Technology of China is supported by the National Natural Science Foundation of China (Grants Nos. 62222512, 12104439, 12134014, and 11974335),  the Anhui Provincial Natural Science Foundation (Grant No. 2208085J03), USTC Research Funds of the Double First-Class Initiative (Grant Nos. YD2030002007 and YD2030002011) and the Fundamental Research Funds for the Central Universities (Grant No. WK2470000035).
The work at Fudan University is  supported by   the National Natural Science Foundation of China (Grant No.~92165109),  National Key Research and Development Program of China (Grant No. 2022YFA1404204), and Shanghai Municipal Science and Technology Major Project (Grant No.~2019SHZDZX01).
\end{acknowledgements}


%

\end{document}


\title{Experimental Demonstration of  Inequivalent Mutually Unbiased Bases: Supplemental Material}
	\maketitle

\section{The Clifford group and restricted Clifford group}\label{app:Clifford orbits}

The Pauli group for one qubit is generated by the following three Pauli matrices:
\begin{equation}
    X = 
    \begin{pmatrix}
        0 & 1 \\ 1 & 0
    \end{pmatrix}, \quad
    Y = 
    \begin{pmatrix}
        0 & -\rmi \\ \rmi & 0
    \end{pmatrix}, \quad
    Z = 
    \begin{pmatrix}
        1 & 0 \\ 0 & -1
    \end{pmatrix}.
\end{equation}
The $n$-qubit Pauli group $\caP_n$ is composed of the $n$-fold tensor products of Pauli matrices and the identity $\{I, X, Y, Z\}^{\otimes n}$ together with the phase factors $\{\pm 1, \pm \rmi\}$, which has order $4^{n+1}$ by construction. In many applications, the overall phases are irrelevant, and it is more convenient to work with the Pauli group $\caP_n$ modulo phase factors, which  is also called the Pauli group (also known as the projective Pauli group) and  is denoted by 
$\bar{\caP}_n$ to distinguish it from $\caP_n$. For the convenience of discussion, the group $\bar{\caP}_n$ can be
identified with the set $\{I, X, Y, Z\}^{\otimes n}$.

The Clifford group $\caC_n$ is the normalizer of the Pauli group $\caP_n$ \cite{Bolt1961I,Bolt1961II,Gottesman1997}. In other words, it is composed of all unitary operators that map the group $\caP_n$ to itself. The Clifford group modulo overall phase factors is also called the Clifford group and is denoted by $\bar{\caC}_n$ in analogy to the  Pauli group $\bar{\caP}_n$. 
The group $\bar{\caC}_n$ can be generated by phase gates $P$, Hadamard gates $H$ for individual qubits and controlled-not (CNOT) gates for all pairs of qubits \cite{Gottesman1997}, where
\begin{equation}
P = 
\begin{pmatrix}
    1 & 0 \\ 0 & \rmi
\end{pmatrix}, \quad
H = \frac{1}{\sqrt{2}}
\begin{pmatrix}
    1 & 1 \\ 1 & -1
\end{pmatrix},  \quad
\mathrm{CNOT} = 
\begin{pmatrix}
    1 & 0 & 0 & 0 \\
    0 & 1 & 0 & 0 \\
    0 & 0 & 0 & 1 \\
    0 & 0 & 1 & 0
\end{pmatrix}.
\end{equation}
The quotient group $\bar{\caC}_n/\bar{\caP}_n$ is isomorphic to the symplectic group $\Sp{2n}{2}$, which  is the group of linear transformations (symplectic matrices) on $\F_2^{2n}$  that preserve the standard symplectic inner product \cite{Bolt1961I,Bolt1961II}.

In this work, we focus on  the two-qubit Clifford group with $n=2$. In this case, the symplectic group $\Sp{4}{2}$ is isomorphic to the symmetric group of six letters and has order 720. Accordingly,  the Clifford group $\bar{\caC}_2$ has order 11520. In addition, we are particularly interested in a special subgroup of the Clifford group, the restricted Clifford group $\bar{\caC}^\rmr_2$ \cite{Zhu2017_3design}, whose quotient over the Pauli group is isomorphic to $\SL{2}{4}$, which in turn is isomorphic to the alternating group $A_5$. Both $\SL{2}{4}$ and $A_5$ have orders 60, so $\bar{\caC}^\rmr_2$ has order 960. To be specific, $\bar{\caC}^\rmr_2$ is generated by the following two unitary operators:
\begin{equation}
    H_2 \rmCNOT_{12}P_1 H_2, \  H_1 P_2 \rmCNOT_{12} H_2,
\end{equation}
where $H_j$ and $P_j$ for $j=1,2$ denote the Hadamard gate and phase gate acting on the $j$th qubit, and $\rmCNOT_{12}$ is the CNOT gate with the first qubit as the control qubit and the second qubit as the target qubit.

\section{Constructing 4-designs from (restricted) Clifford orbits}\label{app:fiducial vector}

Although the Clifford group is a 3-design, but not a 4-design \cite{Zhu2017_3design,Webb2016}, its orbits may form 4-designs if we choose suitable fiducial states \cite{zhu2016clifford}. In the case of two qubits under consideration, we can even construct a 4-design from certain orbit of the restricted Clifford group, which is only a 2-design \cite{Zhu2017_3design}. 

Here we focus on   product fiducial states of the form   $|\psi(\tfrac{1}{\sqrt{3}},\tfrac{1}{\sqrt{3}},\tfrac{1}{\sqrt{3}})\> \otimes |\psi(r_1,r_2,r_3)\>$, 
where each qubit state is characterized by its Bloch vector. According to \rcite{zhu2016clifford}, the  orbit generated by the Clifford group forms a 4-design iff the Bloch vector $(r_1,r_2,r_3)$ satisfies the following condition \cite{zhu2016clifford}:
\begin{align}\label{eq:4designCondition}
r_1^4+r_2^4+r_3^4 = \frac{5}{7}. 
\end{align}
The Bloch vector can also be represented by two spherical angles $\theta, \phi$ via the relation $ (\sin\theta \cos\phi, \sin\theta \sin\phi, \cos\theta) = (r_1,r_2,r_3)$. Then the condition in \eref{eq:4designCondition} can be expressed as follows,
\begin{equation}\label{eq:angle0}
  \cos(4 \phi) = 4 \alpha - 3, \quad 
  \cos(2\theta) = \frac{\alpha-1 \pm \frac{2}{\sqrt{7}} \sqrt{5-2\alpha} }{\alpha+1},\quad \alpha \in [1/2, 1]. 
\end{equation}

In general, the orbit of the restricted Clifofrd group is not a 4-design even if the spherical angles $\theta,\phi$ satisfy \eref{eq:angle0}. Here we are particularly interested in the special value $\alpha=1$, which means
\begin{equation}\label{eq:angle1}
  \phi=0, \quad \theta= \frac{1}{2}\arccos \sqrt{\frac{3}{7}}.
\end{equation}
The corresponding Bloch vector reads
\begin{equation}
    (r_1,r_2,r_3) = \left(-\sqrt{\frac{1}{2}-\frac{1}{2}\sqrt{\frac{3}{7}}},\; 0, \;\sqrt{\frac{1}{2}+\frac{1}{2}\sqrt{\frac{3}{7}}}\, \right).
\end{equation}
Up to overall phase factors, the stabilizer of the fiducial state within the Clifford group consists of the three unitary operators $I \otimes I, HPZ \otimes I, HPHPX \otimes I$, and  the orbit length is 3840. By contrast, the stabilizer within the restricted Clifford group is trivial, and the orbit length is 960. Surprisingly, the orbit generated by the restricted Clifford group also forms a 4-design, although the restricted Clifford group is only a 2-design \cite{Zhu2017_3design}.
This result can be verified by direct calculation based on the $t$th frame potential with $t=4$ [see \eref{eq:FPsupp} below].
This 4-design is appealing to experimental realization because it has a small
 cardinality.

\section{Numerical generation of 4-designs}\label{app:numerical 4-design}

The $t$th \emph{frame potential} of a set of $K$ states $\{|\psi_j\>\}_j$ is defined as  \cite{Renes2004, Zauner2011, Scott2006}
\begin{equation}\label{eq:FPsupp}
    \Phi_t(\{|\psi_j \>\}_j) = \frac{1}{K^2} \sum_{j,k}| \< \psi_j|\psi_k\>|^{2t}.
\end{equation}
It satisfies the inequality
\begin{align}
    \Phi_t(\{|\psi_j \>\}_j) \ge \frac{1}{D_t},
\end{align}
and the lower bound is saturated iff the set $\{|\psi_j\>\}_j$ forms a $t$-design. This result is particularly convenient to determining whether a set forms a $t$-design or not. Moreover, given an initial set of states, better and better approximate $t$-designs can be constructed by minimizing the $t$th frame potential. A concrete optimization algorithm is provided in \rscite{HUGHES202184,tightframes}. By virtue of  this algorithm (with iteration number $k=2\times 10^6$ and  random seed number $s=10^5$) we generated an approximate numerical 4-design composed of 200 states in dimension 4. The corresponding frame potential is 0.02859873, which is 0.1\% larger than the lower bound $1/35=0.02857143$. The ratio between the smallest and largest eigenvalues (within the symmetric subspace) of the moment operator $\sum_j (|\psi_j\>\<\psi_j|)^{\otimes 4}$ is about 0.90. Although this figure of merit is not so close to the ideal value of 1, the numerical 4-design is accurate enough for our purpose 
according to the following analysis; cf. Figs. 3, 4 and Table II in the main text as well as Sec.~S4 in this Supplemental Material.

\begin{figure}[bt]
    \centering		      \includegraphics[width=0.6\textwidth]{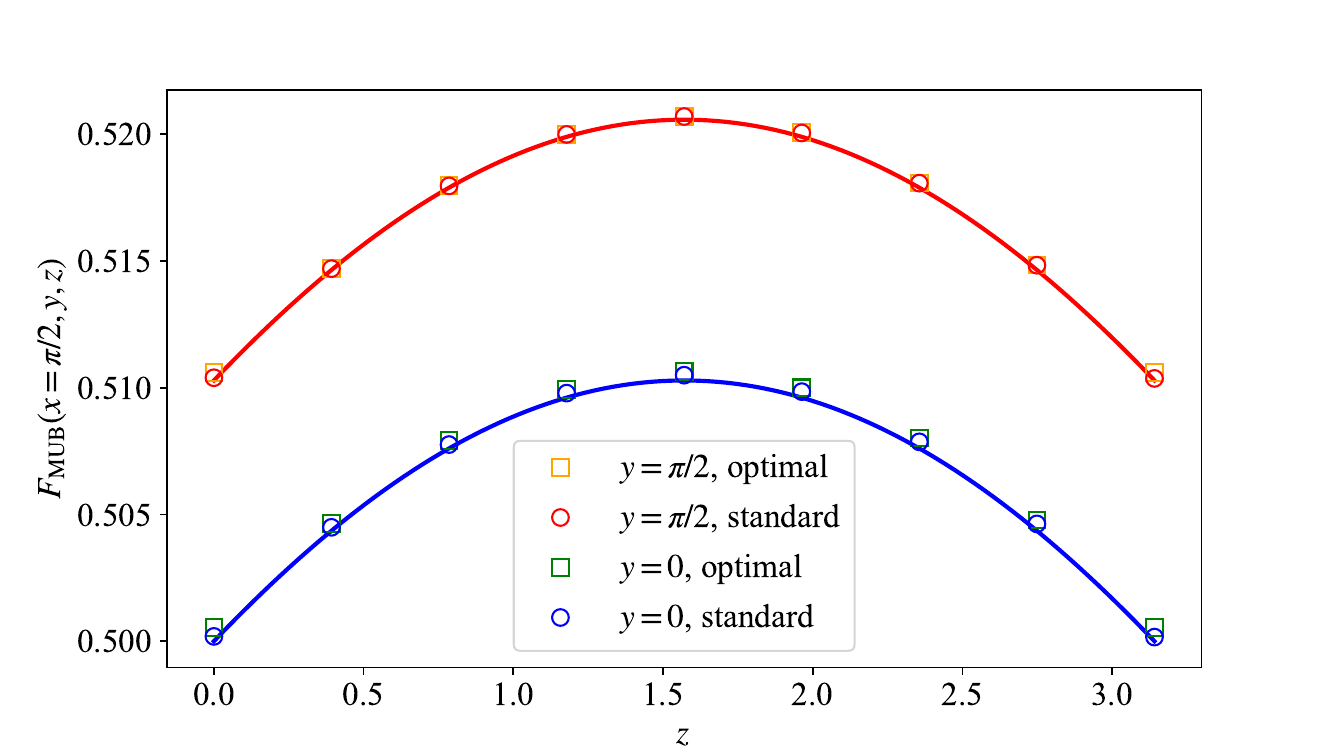}  
    \caption{ \label{fig:numP}  The optimal and standard estimation fidelities of 18 triples of MUB based on the numerical 4-design in comparison with the ideal values (solid lines). 
    }
\end{figure} 

If the state ensemble forms an ideal 4-design, then the estimation fidelity of the POVM $\scrA=\{A_j\}_j$ is determined by Eqs.~(1) and (2) in the main text. Since the numerical 4-design is not ideal, to determine the optimal estimation fidelity $Q(A_j)$ in Eq.~(2) should be replaced by 
\begin{equation}
    Q'(A_j) := (N+1)!\tr_{1,\dots,N} [P'_{N+1} (A_j \otimes \id)],
\end{equation}
where
\begin{equation}
P'_{N+1}=\frac{D_{N+1}}{K} \sum_j (|\psi_j\>\<\psi_j|)^{\otimes 4}.
\end{equation}
Accordingly, the optimal estimator should be supported in the eigenspace of $Q'(A_j)$ with the largest eigenvalue. \Fref{fig:numP} shows the optimal estimation fidelities of 18 triples of MUB based on the numerical 4-design in comparison with the ideal values based on an exact 4-design.  The maximal deviation from the ideal estimation fidelity is only 0.00053, which is much smaller than the maximal difference between estimation fidelities of inequivalent MUB. If instead we use the standard  estimator  supported in the eigenspace of $Q(A_j)$ with the largest eigenvalue (which is suboptimal), then the estimation fidelity that can be achieved is slightly smaller, but still close to the ideal value, as illustrated in \fref{fig:numP}. In this case, the maximal deviation from the ideal estimation fidelity is only 0.00027. These results show that the numerical 4-design is accurate enough for distinguishing inequivalent MUB.

\section{Experimental implementation of the target projective measurements}\label{app:measurement}

\begin{figure}[bp]
\begin{minipage}[b]{\linewidth}
\centering
\hspace{1.2ex}\includegraphics[width=0.85\textwidth]{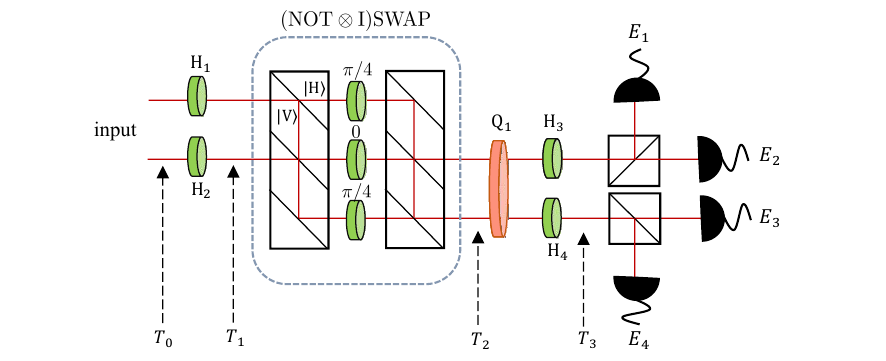}

\end{minipage}
\begin{minipage}[b]{\linewidth}
\centering
\vspace{2ex}

\begin{tabular}{p{2.5cm}<{\centering} | p{1.5cm}<{\centering} p{1.5cm}<{\centering} p{1.5cm}<{\centering} p{2cm}<{\centering} p{2cm}<{\centering}}
\hline \hline
measurement & $\mathrm{H_1}$ & $\mathrm{H_2}$ & $\mathrm{Q_1}$ & $\mathrm{H_3}$ & $\mathrm{H_4}$ \\  \hline
$\scrA$ & 0 & 0 & 0 & 0 & 0 \\
$\scrB$ & $\pi/8$ & $\pi/8$ & $\pi/4$ & $\pi/8$ & $x/4+\pi/4$\\
$\scrC$ & $3\pi/8$ & $\pi/8$ & $\pi/4$ & $z/4+\pi/8$ & $y/4+\pi/8$\\
\hline \hline
\end{tabular}
\end{minipage}
\caption{\label{fig:measurement} Experimental realization of the projective measurements $\scrA$, $\scrB$, and $\scrC$ associated with the triple of MUB corresponding to the parameter point $(x,y,z)$. The rotation angles of the relevant wave plates outside the dashed box are specified in the embedded table. }
\end{figure}

In this section, we show how the measurement module in Fig.1 in the maintext (also shown in \fref{fig:measurement}) performs the target projective measurements $\scrA$, $\scrB$, and $\scrC$ associated with the triple of MUB corresponding to the parameter point $(x,y,z)$. The polarization (horizontal $\ket{H}$ and vertical $\ket{V}$) and the path (up $\ket{\uparrow}$ and down $\ket{\downarrow}$) degrees of freedom of a single photon act as a ququad. The submodule in the dashed box composed of two BDs and three HWPs set at $45^\circ$, $0^\circ$, and $45^\circ$ implements the unitary transformation $\mathrm{(NOT\otimes I)SWAP}$, where $\mathrm{NOT}$, $\mathrm{I}$, and $\mathrm{SWAP}$ refer to the NOT gate,  identity gate, and  SWAP gate between the polarization and path qubits. The rotation angles of the other wave plates are specified in the embedded table. To verify the validity of the setup,  the evolution of the basis states of $\scrA$, $\scrB$, and $\scrC$, i.e., $\ket{\alpha_j}$, $\ket{\beta_k}$, and $\ket{\gamma_l}$ for $j, k, l  \in \{1,2,3,4\}$ is summarized in \tref{tab:evolution}.
 
\begin{table*}[htbp]
\renewcommand{\arraystretch}{1.75}\centering
\caption{\label{tab:evolution}  The evolution of the basis states associated with the projective measurements $\scrA$, $\scrB$, and $\scrC$ from time $T_0$ to time $T_3$ as indicated in \fref{fig:measurement} and the corresponding measurement outcomes.} 
	\begin{tabular}{p{1.2cm}<{\centering} | p{6cm}<{\centering} p{3.6cm}<{\centering} p{3.8cm}<{\centering} p{1.2cm}<{\centering} p{1.2cm}<{\centering} }  
		\hline \hline 
  
		input  &  $T_0$ & $T_1$ & $T_2$ & $T_3$ & outcome\\
		\hline
  
		$\ket{\alpha_1}$ & $\ket{H}\ket{\uparrow}$ & $\ket{H}\ket{\uparrow}$  & $\ket{V}\ket{\uparrow}$ & $\ket{V}\ket{\uparrow}$ & $E_1$  \\

		$\ket{\alpha_2}$ & $\ket{H}\ket{\downarrow}$ & $\ket{H}\ket{\downarrow}$  & $\ket{H}\ket{\uparrow}$ & $\ket{H}\ket{\uparrow}$ & $E_2$\\
		
		$\ket{\alpha_3}$ & $\ket{V}\ket{\uparrow}$ &  $\ket{V}\ket{\uparrow}$ & $\ket{V}\ket{\downarrow}$ & $\ket{V}\ket{\downarrow}$ & $E_4$\\
		
		$\ket{\alpha_4}$ & $\ket{V}\ket{\downarrow}$  & $\ket{V}\ket{\downarrow}$  & $\ket{H}\ket{\downarrow}$ & $\ket{H}\ket{\downarrow}$ & $E_3$\\ 
            \hline

		$\ket{\beta_1}$ & $\frac{1}{2}(\ket{H}+\ket{V})(\ket{\uparrow}+\ket{\downarrow})$ & $\frac{1}{\sqrt{2}}\ket{H}(\ket{\uparrow}+\ket{\downarrow})$ & $\frac{1}{\sqrt{2}}(\ket{H}+\ket{V})\ket{\uparrow}$ & $\ket{H}\ket{\uparrow}$ & $E_2$ \\
		 
		$\ket{\beta_2}$ & $\frac{1}{2}(\ket{H}-\ket{V})(\ket{\uparrow}+\rmi\rme^{\rmi x}\ket{\downarrow})$ & $\frac{1}{\sqrt{2}}\ket{V}(\ket{\uparrow}+\rmi\rme^{\rmi x}\ket{\downarrow})$ & $\frac{1}{\sqrt{2}}(\rmi\rme^{\rmi x}\ket{H}+\ket{V})\ket{\downarrow}$ & $\ket{H}\ket{\downarrow}$ & $E_3$\\
		 
		$\ket{\beta_3}$ & $\frac{1}{2}(\ket{H}+\ket{V})(\ket{\uparrow}-\ket{\downarrow})$ & $\frac{1}{\sqrt{2}}\ket{H}(\ket{\uparrow}-\ket{\downarrow})$ & $\frac{1}{\sqrt{2}}(-\ket{H}+\ket{V})\ket{\uparrow}$ & $\ket{V}\ket{\uparrow}$ & $E_1$\\
		 
		$\ket{\beta_4}$ & $\frac{1}{2}(\ket{H}-\ket{V})(\ket{\uparrow}-\rmi \rme^{\rmi x}\ket{\downarrow})$ & $\frac{1}{\sqrt{2}}\ket{V}(\ket{\uparrow}-\rmi \rme^{\rmi x}\ket{\downarrow})$ & $\frac{1}{\sqrt{2}}(-\rmi \rme^{\rmi x}\ket{H}+\ket{V})\ket{\downarrow}$ & $\ket{V}\ket{\downarrow}$ & $E_4$ \\
            \hline

		$\ket{\gamma_1}$ & $\frac{1}{2}[(\ket{H}+\ket{V})\ket{\uparrow}-\rme^{\rmi y}(\ket{H}-\ket{V})\ket{\downarrow}]$ & $\frac{1}{\sqrt{2}}\ket{V}(\ket{\uparrow}-\rme^{\rmi y}\ket{\downarrow})$  & $\frac{1}{\sqrt{2}}(-\rme^{\rmi y}\ket{H}+\ket{V})\ket{\downarrow}$ & $\ket{V}\ket{\downarrow}$  & $E_4$\\

		$\ket{\gamma_2}$ & $\frac{1}{2}[(\ket{H}-\ket{V})\ket{\uparrow}+\rme^{\rmi z}(\ket{H}+\ket{V})\ket{\downarrow}]$ & $\frac{1}{\sqrt{2}}\ket{H}(\ket{\uparrow}-\rme^{\rmi z}\ket{\downarrow})$ & $\frac{1}{\sqrt{2}}(\rme^{\rmi z}\ket{H}-\ket{V})\ket{\uparrow}$ & $\ket{V}\ket{\uparrow}$ & $E_1$\\

		$\ket{\gamma_3}$ & $\frac{1}{2}[(\ket{H}+\ket{V})\ket{\uparrow}+\rme^{\rmi y}(\ket{H}-\ket{V})\ket{\downarrow}]$ &  $\frac{1}{\sqrt{2}}\ket{V}(\ket{\uparrow}+\rme^{\rmi y}\ket{\downarrow})$ & $\frac{1}{\sqrt{2}}(\rme^{\rmi y}\ket{H}+\ket{V})\ket{\downarrow}$ & $\ket{H}\ket{\downarrow}$ & $E_3$ \\

		$\ket{\gamma_4}$ & $\frac{1}{2}[(\ket{H}-\ket{V})\ket{\uparrow}-\rme^{\rmi z}(\ket{H}+\ket{V})\ket{\downarrow}]$  & $\frac{1}{\sqrt{2}}\ket{H}(\ket{\uparrow}+\rme^{\rmi z}\ket{\downarrow})$ & $\frac{1}{\sqrt{2}}(\rme^{\rmi z}\ket{H}+\ket{V})\ket{\uparrow}$ & $\ket{H}\ket{\uparrow}$ & $E_2$\\
		\hline \hline 
	\end{tabular}
\end{table*}

\section{Estimation fidelities of equivalent MUB}
\begin{figure}[htbp]
    \centering		     
    \includegraphics[width=0.7\textwidth]{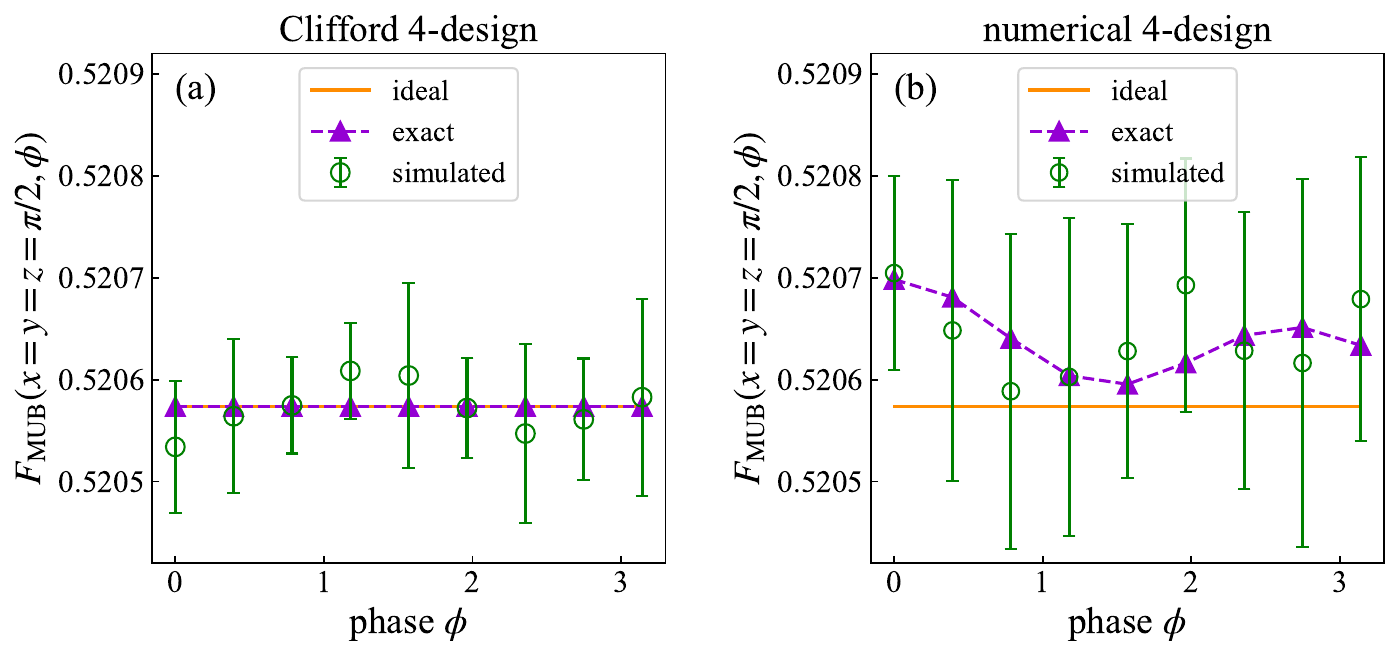}  
    \caption{ \label{fig:equi_MUB}  
    Exact and simulated three-copy estimation fidelities of equivalent MUB constructed by applying the controlled-phase gate $\diag(1,1,1,\rme^{\rmi\phi})$ to the MUB corresponding to the parameter point $x=y=z=\pi/2$. In numerical simulation, each measurement is repeated $10000\times10$ times. The state ensemble corresponds to the Clifford 4-design in plot (a) and numerical 4-design in  plot (b).
    }
\end{figure}

Here we provide some auxiliary results on the three-copy estimation fidelities of equivalent MUB in dimension 4. We start with the triple of MUB corresponding to the parameter point $x=y=z=\pi/2$ and apply the  controlled phase gate $\diag(1,1,1,\rme^{\rmi\phi})$ on the three bases simultaneously. If the state ensemble forms an ideal 4-design, which is the case for the Clifford 4-design, then the exact estimation probability  is equal to $(46+5\sqrt{3})/105 \simeq 0.52057$, which is invariant under any unitary transformation of the MUB, as illustrated in plot (a) in \fref{fig:equi_MUB}. This plot also shows the estimation fidelity together with the standard deviation determined by numerical simulation with each measurement repeated $10000\times 10$ times. The statistical fluctuation causes a small variation of the estimation fidelity with $\phi$, whose order of magnitude is around $10^{-4}$.

If the Clifford 4-design is replaced by the numerical 4-design, then even the exact estimation fidelity may depend on the phase as illustrated in plot (b) in \fref{fig:equi_MUB}. Fortunately, this variation is even smaller than the one induced by statistical fluctuation and is negligible compared with the maximal difference between estimation fidelities of inequivalent MUB.

\begin{table*}[htbp]
\renewcommand{\arraystretch}{1.5}
\centering
\caption{\label{tab:equiMUB}
Exact and simulated three-copy estimation fidelities of equivalent MUB constructed by applying 10000 random unitaries to the MUB corresponding to the parameter point $x=y=z=\pi/2$. Also shown are the standard deviations and maximal deviations over the random unitaries.  In numerical simulation, each measurement is repeated $10000\times10$ times. The state ensemble corresponds to either  the Clifford 4-design or the numerical 4-design.} 
\begin{tabular}{p{1.5cm}<{\centering} | p{1.5cm}<{\centering} p{1.5cm}<{\centering} p{1.5cm}<{\centering} p{1.4cm}<{\centering} p{1.5cm}<{\centering} |p{1.5cm}<{\centering} p{1.5cm}<{\centering} p{1.5cm}<{\centering} p{1.4cm}<{\centering} p{1.5cm}<{\centering}  }  \hline \hline
\multirow{2}{*}{\textbf{4-design}} & \multicolumn{5}{c|}{\textbf{exact estimation fidelity}} & \multicolumn{5}{c}{\textbf{simulated estimation fidelity}} \\ \cline{2-11}
  &  maximal & minimal & average & std. & maximal deviation & maximal & minimal & average & std. & maximal deviation \\ \hline
Clifford & 0.52057 & 0.52057 & 0.52057 & 0 & 0 & 0.52080 & 0.52025 & 0.52057 & 0.00006 & 0.00032  \\ 
numerical  &  0.52083 & 0.52030 & 0.52057 & 0.00007 & 0.00027 & 0.52123 & 0.51995 & 0.52057 & 0.00016 & 0.00066 \\ \hline \hline
\end{tabular}
\end{table*}

\begin{figure}[htbp]
    \centering		     
    \includegraphics[width=0.6\textwidth]{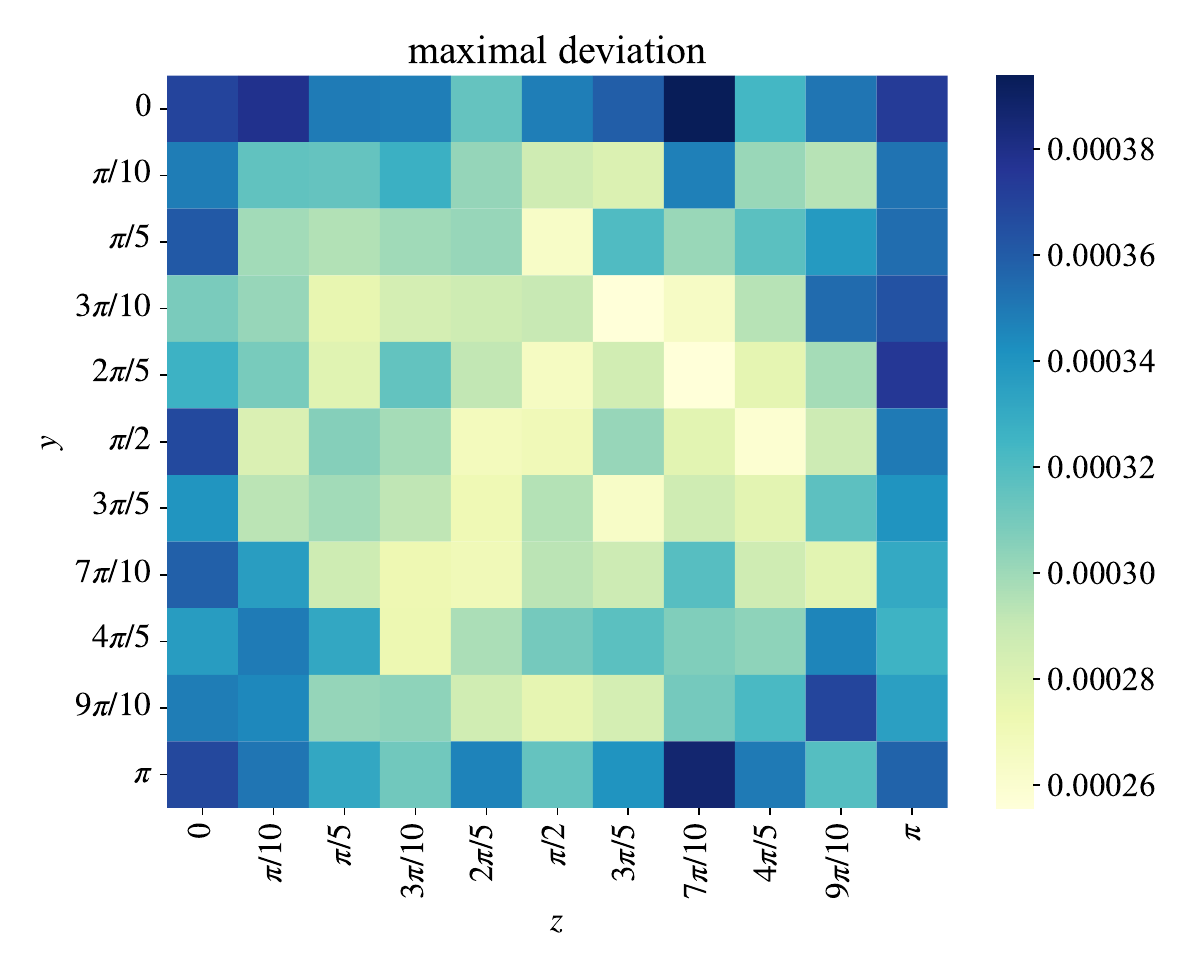}  
    \caption{ \label{fig:equiU_theodiff}  
    Maximal deviations of the exact three-copy estimation fidelities (from the ideal values) of equivalent MUB constructed by applying 10000 random unitaries to MUB corresponding to the $y-z$ plane with $x=\pi/2$. The state ensemble corresponds to  the numerical 4-design.
    }
\end{figure}

Next, we consider the impact of random unitary transformations instead of the controlled phase gate on the triple of MUB corresponding to the parameter point $x=y=z=\pi/2$. In total 10000 random unitaries are generated and applied. \Tref{tab:equiMUB} summarizes the maximal, minimal, and average estimation fidelities together with the standard deviations and maximal deviations based on the Clifford 4-design and numerical 4-design, respectively. Again, the deviations are much smaller than the maximal difference between estimation fidelities of inequivalent MUB. For the numerical 4-design,
\fref{fig:equiU_theodiff} further shows 
the maximal deviations between the exact estimation fidelities and the ideal estimation fidelities for MUB corresponding to the $y-z$ plane with $x=\pi/2$. The order of magnitude is similar to the counterpart in \tref{tab:equiMUB}.
	
\section{Estimation fidelities 
based on random subsets of the Clifford 4-design}
\begin{figure}[tb]
    \centering		     
\includegraphics[width=0.7\textwidth]{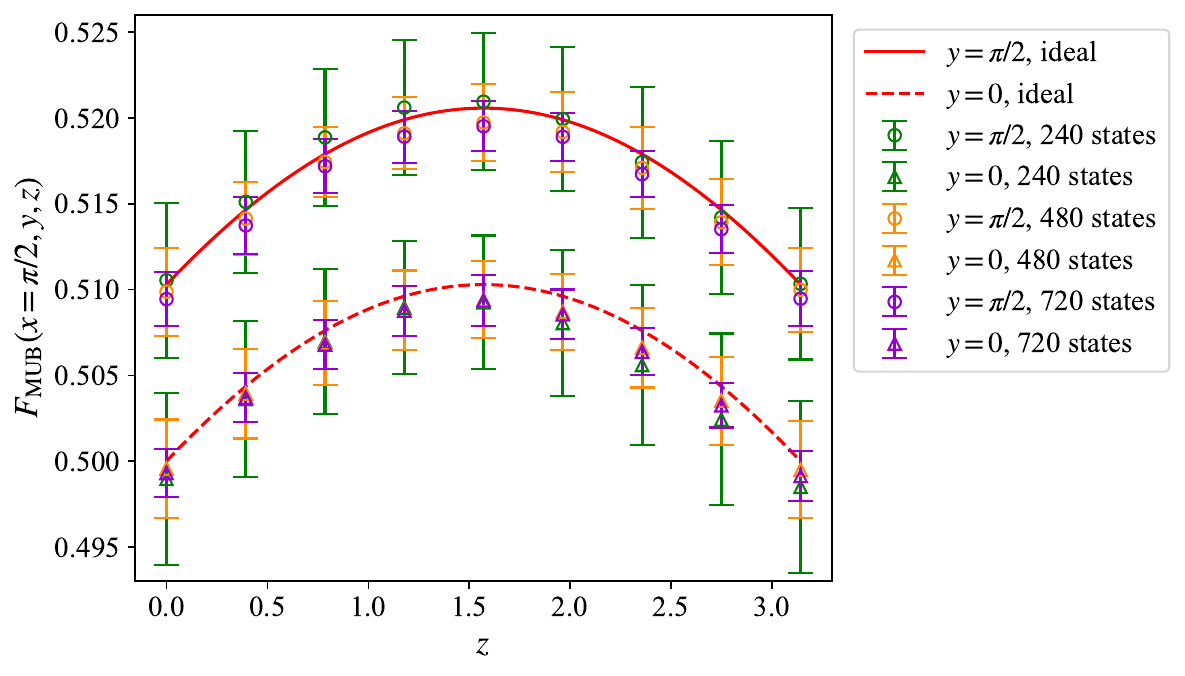}
    \caption{\label{fig:randCli} 
    Estimation fidelities  $F_\mathrm{MUB}(x,y,z)$ 
   based on random subsets of the Clifford 4-design with 240, 480, and 720 states, respectively. 
 Each data point is an average over 30 selections sampled from the experimental data. The error bar indicates the standard deviation over the 30 selections. The ideal estimation fidelity is also plotted for comparison.}
\end{figure}

In the maintext we determined the estimation fidelities of triples of MUB by averaging over all states in the Clifford 4-design. Here we consider alternative approaches by averaging over random subsets of the Clifford 4-design. To this end,
we randomly select $K$ distinct states from the Clifford 4-design and compute the estimation fidelity $F_\mathrm{MUB}(x, y,z)$ for each triple of MUB by averaging over the experimental data (the average frequency of each outcome determined by $10000\times 10$ measurements)  corresponding to the $K$ states. This procedure is repeated 30 times to determine the mean estimation fidelity and the standard deviation. 
The main results for $K=240, 480, 720$ are illustrated in \fref{fig:randCli} and summarized in \tref{tab:randClif}. For all three choices, the mean estimation fidelities thus determined are close to the ideal value based on the Clifford 4-design. Moreover, the standard deviation tends to decrease monotonically as $K$ increases as expected.

\begin{table*}[tbp]
        \renewcommand{\arraystretch}{1.5}
	\centering
	\caption{\label{tab:randClif} 
     Estimation fidelities  $F_\mathrm{MUB}(x,y,z)$ 
   based on random subsets of the Clifford 4-design with 240, 480, and 720 states, respectively. 
 Each data point is an average over 30 selections sampled from the experimental data. The error bar indicates the standard deviation over the 30 selections.  The ideal estimation fidelity and experimental estimation fidelity based on all 960 states of the Clifford 4-design are also shown for comparison.
 }
	\begin{tabular}{p{1cm}<{\centering} |p{0.8cm}<{\centering} | p{1.6cm}<{\centering} p{1.6cm}<{\centering} p{1.6cm}<{\centering} p{1.6cm}<{\centering} p{1.6cm}<{\centering} p{1.6cm}<{\centering} p{1.6cm}<{\centering} p{1.6cm}<{\centering} p{1.6cm}<{\centering}  }  
		\hline \hline 
            \multicolumn{2}{c|}{$z$}   &  0 & $\frac{\pi}{8}$ & $\frac{\pi}{4}$ & $\frac{3\pi}{8}$ & $\frac{\pi}{2}$ & $\frac{5\pi}{8}$ & $\frac{3\pi}{4}$ & $\frac{7\pi}{8}$ & $\pi$\\
		\hline
		& 240 & 0.5105(45)& 0.5151(41)& 0.5189(40)& 0.5206(39) & 0.5209(40)& 0.5199(42)& 0.5174(44)& 0.5142(45)& 0.5103(44)\\
		 
		$x=\frac{\pi}{2}$& 480 & 0.5099(26)& 0.5141(21)& 0.5174(20)& 0.5191(21)& 0.5197(22)& 0.5192(23)& 0.5171(24)& 0.5139(25)&	0.5100(24)  \\
		
		$y=\frac{\pi}{2}$& 720 & 0.5094(16)& 0.5137(16)& 0.5172(16)& 0.5189(15)& 0.5195(15)& 0.5189(14)& 0.5167(13)& 0.5135(14)& 0.5095(16)  \\

		& 960 & 0.5096& 0.5137& 0.5172& 0.5190& 0.5197& 0.5192& 0.5171& 0.5139& 0.5096  \\
  
  		&ideal &0.5103& 0.5146& 0.5179& 0.5199& 0.5206& 0.5199& 0.5179& 0.5146& 0.5103\\

		\hline 
		& 240 & 0.4990(50)& 0.5036(46)& 0.5070(42)& 0.5090(39)& 0.5093(39)&0.5081(42)& 0.5056(47)& 0.5024(50)& 0.4985(50)\\
		
		$x=\frac{\pi}{2}$& 480 & 0.4996(29)& 0.5039(26)& 0.5069(24)& 0.5088(23)& 0.5094(22)& 0.5087(22)& 0.5066(23)& 0.5035(26)& 0.4995(28)\\
		
		$y=0$& 720 & 0.4993(14)& 0.5037(14)& 0.5068(14)& 0.5088(15)& 0.5094(15)& 0.5086(14)& 0.5064(14)& 0.5033(13)& 0.4991(15)\\
  
            & 960 & 0.4994& 0.5036& 0.5068& 0.5088& 0.5095& 0.5088& 0.5067& 0.5036& 0.4992 \\
            
		&ideal &0.5000   & 0.5044& 0.5076& 0.5096& 0.5103& 0.5096& 0.5076& 0.5044&	0.5000  \\
            
		\hline \hline 
	\end{tabular}
    \end{table*}